\documentclass[a4paper,11pt]{article}
\pdfoutput=1 
\usepackage{jinstpub} 
\usepackage{lineno}

\title{\boldmath Timing Optimisation and Analysis in the Design of 3D silicon sensors: the TCoDe Simulator}

\author[a]{Angelo Loi,}
\author[a]{Andrea Contu,}
\author[a,1]{Adriano Lai,}\note{Corresponding author.}

\affiliation[a]{Istituto Nazionale Fisica Nucleare, Sezione di Cagliari, Cagliari, Italy}

\emailAdd{adriano.lai@ca.infn.it}

\abstract{
Solid state sensors having timing capabilities are becoming a mandatory requirement in particle tracking techniques of future experiments at colliders. Within this context, silicon sensors with 3D structure provide an interesting solution, due to their intrinsic fast response and radiation resistance. The performance of such devices is strictly dependent on the geometric structure of the electrodes and can be considerably optimised by design using suitable tools to accurately model the sensor behaviour. This paper illustrates the development, performance and use of the TCoDe simulator, specifically dedicated to the fast simulation of carrier transportation phenomena in solid state sensors. Some examples of its effectiveness in the design and analysis of 3D sensors are also given.
}

\arxivnumber{2008.10730v2} 

\begin{document}
\maketitle
\flushbottom

\section{Introduction}
\label{sec:intro}

Tracking systems of the next-to-come high-energy physics experiments at colliders, and vertex detectors in particular, will cope with an extremely large number of tracks per collision. Such experimental challenge could greatly benefit from a precise measurement of the time coordinate of the tracking hits. This solution has been pioneered by the NA62 experiment, with resolutions around 150~ps~ per hit \cite{gtk}, and has already been adopted for the tracking systems of the Phase-II Upgrades of the ATLAS and CMS experiments at the Large Hadron Collider (LHC), where dedicated \emph{Timing Layers}~\cite{ATLAS-TL, CMS-TL}, aiming at resolutions around 30~ps per track are currently under development. In the high-luminosity Upgrade-2 of the LHCb experiment~\cite{LHCb-PII-Physics,LHCb-PII-EoI}, planned for the Run 5 of the LHC, this requirement is even more demanding, because, in order to ensure satisfactory track reconstruction efficiency, the timing information is needed at the single pixel level of the vertex detector. Similar and more stringent conditions are expected for the next generation of colliders, that is those of the so-called \emph{Future Circular Collider} family ~\cite{FCC,RiassuntoEUStrat,CEPC}.

This experimental approach involves the detector system as a whole (from sensors to the front-end electronics to the processing stage). The TimeSPOT Project aims at developing a full tracker demonstrator of reduced size (a few thousand channels and 4 to 5 tracking planes), in order to conceive the single system constituents while taking into account the inter-relationship and inter-connections of the single parts, from the sensor to the reconstruction algorithms~\cite{timespot}. The development of very radiation-hard and high-resolution sensors is a crucial starting step in the design phase of such systems. Required time resolutions must be at least 50 ps per hit~\cite{FTDR}, which have to be maintained up to high particle fluences (larger than $10^{16}$~${\rm n}_{\rm eq}$/cm$^2$ (1~MeV neutron equivalent/cm$^2$). 

The TimeSPOT collaboration has already developed dedicated sensors for timing, which show excellent time resolution, in the range of 15-20 ps~\cite{PSIres}. 
This paper illustrates the conception and operation of a custom simulation tool, named TCoDe (TimeSPOT Code for Detector simulation), which has played a crucial role in the design and detailed simulation of the sensors produced, making it possible to fully characterise the sensor models with very high statistics within reasonable processing times. For similar reasons, TCoDe is also very useful in a detailed study and interpretation of the measured sensor behavior.
The paper is organised as follows. Section~\ref{sec:sss_timing} recalls some basic principles about time resolution in silicon sensors. Section~\ref{sec:3D_design} illustrates the relevant design parameters in defining the sensor simulation model and the main steps in sensor design. Section~\ref{sec:3D_optim} describes the conception and operation of the TCoDe software, while section~\ref{sec:time_behave} illustrates some examples of its practical usage. Finally, section~\ref{sec:conclusion} draws our conclusions.

\section{Solid state sensors for timing}
\label{sec:sss_timing}

In a generic silicon sensor, it is possible to summarise the main contributions to time resolution $\sigma_{\rm t}$ by means of the formula:

\begin{equation}\label{eq1}
   \sigma_{\rm t} = \sqrt {\sigma^2_{\rm ej} + \sigma^2_{\rm tw}  + \sigma^2_{\rm un} + \sigma^2_{\rm dr} + \sigma^2_{\rm TDC}} \, ,
\end{equation}

where $\sigma_{\rm ej}$ is the electronic jitter, depending on electronic noise and pre-amplifier speed; $\sigma_{\rm tw}$ is due to the time-walk fluctuations, depending on different times of discriminator threshold crossing of signals with same shape and different amplitudes; $\sigma_{\rm dr}$ is due to the time fluctuations due to deposits of secondary delta-rays during the ionisation process; $\sigma_{\rm TDC}$ depends only on the Time-to-Digital-Converter resolution. The term $\sigma_{\rm un}$ (field unevenness) is strictly related to the uniformity in the weighting field and drift velocity across the sensor volume, as established by the Ramo theorem~\cite{Ramo}.

Among the contributions in eq.~\ref{eq1}, the term $\sigma_{\rm ej}$ depends both on sensor (sensor noise) and electronics, the terms $\sigma_{\rm tw}$ and $\sigma_{\rm TDC}$ are independent of the sensor, while $\sigma_{\rm dr}$ and $\sigma_{\rm un}$ depend only on the sensor geometry and therefore can be optimised by design for fast timing.

\subsection{Characteristics of 3D silicon sensors and timing}
\label{subsec:3D_char}

Silicon sensors based on 3D electrode geometry were proposed more than two decades ago by S. Parker~\cite{3D, Parker}. Unlike their planar counterpart, the electrode structure of 3D silicon sensors is developed in vertical direction, orthogonal to their surface. At the expenses of a more complex production process, this structure bears a number of important advantages. 
As the sensor thickness is made independent of the inter-electrode distance, the charge carrier drift length can be made very short (around 20-30~$\mu$m or even less), and the induced current signals very fast, while preserving the amount of charge deposited by ionisation inside the sensor volume, and therefore the signal amplitude.

The vertical electrodes roughly follow the path of the ionising particle to be detected. This makes the effect of delta rays ($\sigma_{\rm dr}$ in eq.~\ref{eq1}) on time fluctuations negligible, improving time resolution. It is worth recalling here that delta rays, sometimes referred to as \emph{Landau fluctuations}, pose at approximately 20 ps the intrinsic limit of time performance for other silicon sensors based on timing--optimised planar technologies~\cite{SADROZINSKI}.

The short inter-electrode distance has also a  beneficial effect on radiation hardness, as it reduces the trapping probability of the charge carriers while they travel towards the collecting electrodes. Indeed, 3D silicon sensors show an unmatched radiation resistance, being successfully tested up to $3\times10^{17}~{\rm n}_{\rm eq}/cm^2$~\cite{3Dradhard}.

In 3D silicon sensors, being the $\sigma_{\rm dr}$ term naturally at its minimum, the term $\sigma_{\rm un}$ is crucial for the improvement of timing performance. It depends strongly on the geometrical structure of the sensor sensitive volume. Once the velocity saturation regime is reached, the more uniform the weighting field, the more uniform the shape of the signals and consequently the smaller the dispersion in the time of arrival of the signals. On the other hand, the sensor contribution to the noise, affecting the term $\sigma_{\rm ej}$, also depends on the geometry of the sensor volume through its capacitance and therefore must be carefully considered during the design optimisation process. 

\section{Analysis of static properties and 3D silicon sensor design}
\label{sec:3D_design}

Once the specific technology and sensor pitch are chosen, the timing optimisation and detailed design of 3D silicon sensors starts with an accurate definition of the pixel geometrical structure. In this first stage, the so-called \emph{static properties} of the pixel are studied and defined. Such physical properties fix the operating conditions of the sensor. They are the electric field, which decides the carrier velocities, and the weighting field. These \emph{static properties} strictly depend on the electrode structure and on the bias voltage and correspond to well defined value maps across the whole sensor volume. They establish the playground where the carrier dynamics and signal generation are then simulated and characterised.

The study of the \emph{static properties} has been performed with the Synopsys Sentaurus TCAD~\cite{tcad} software.

\subsection{Technology and starting design criteria}

\begin{figure}[h!]
    \centering
    \includegraphics[width=0.75\textwidth]{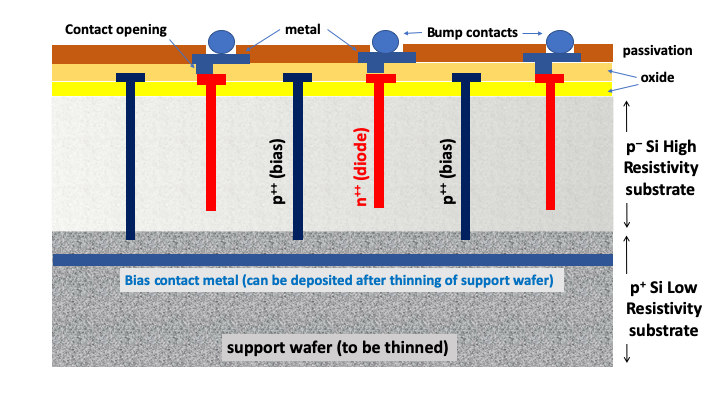}
    \caption{\footnotesize{Schematic of the internal structure of a 3D silicon sensor in Single-Sided technology.}}
    \label{fig:3dstruct}
    \end{figure}

The designed 3D sensors which have been fabricated using a Single-Sided (Si-Si) fabrication approach and the Deep Reactive Ion Etching (DRIE) process~\cite{Bosch}. The Si-Si, as opposed to the double-sided technology used for the ATLAS-IBL 3D sensors~\cite{ATLAS-IBL, ATLAS-IBL2}, allows using a thick support wafer directly connected to the high-resistivity device layer, which improves the mechanical stability of wafers during fabrication and reduces wafer bowing due to mechanical stresses during production. In this way sensitive layers as thin as 150 $\mu$m and electrodes having diameter (column shapes) or width (trench shapes) of about 5 $\mu$m can be fabricated. These values have been adopted as reference minimum sizes in our design. Figure~\ref{fig:3dstruct} shows a schematic of the Si-Si structure. On the other hand, the 150 $\mu$m sensitive silicon depth, still provides a Most Probable Value (MPV) of 2 fC charge deposit, which appears sufficient for our purposes, as demonstrated in the following sections.
The pixel size is kept on a pixel pitch of 55 $\mu$m in order to be compatible with the TIMEPIX ASIC family~\cite{Timepix4}. This size was adopted uniquely for practical reasons, namely to have alternative means to test the pixel matrices produced. 
    
\begin{figure}[h!]
    \centering
    \includegraphics[width=1\textwidth]{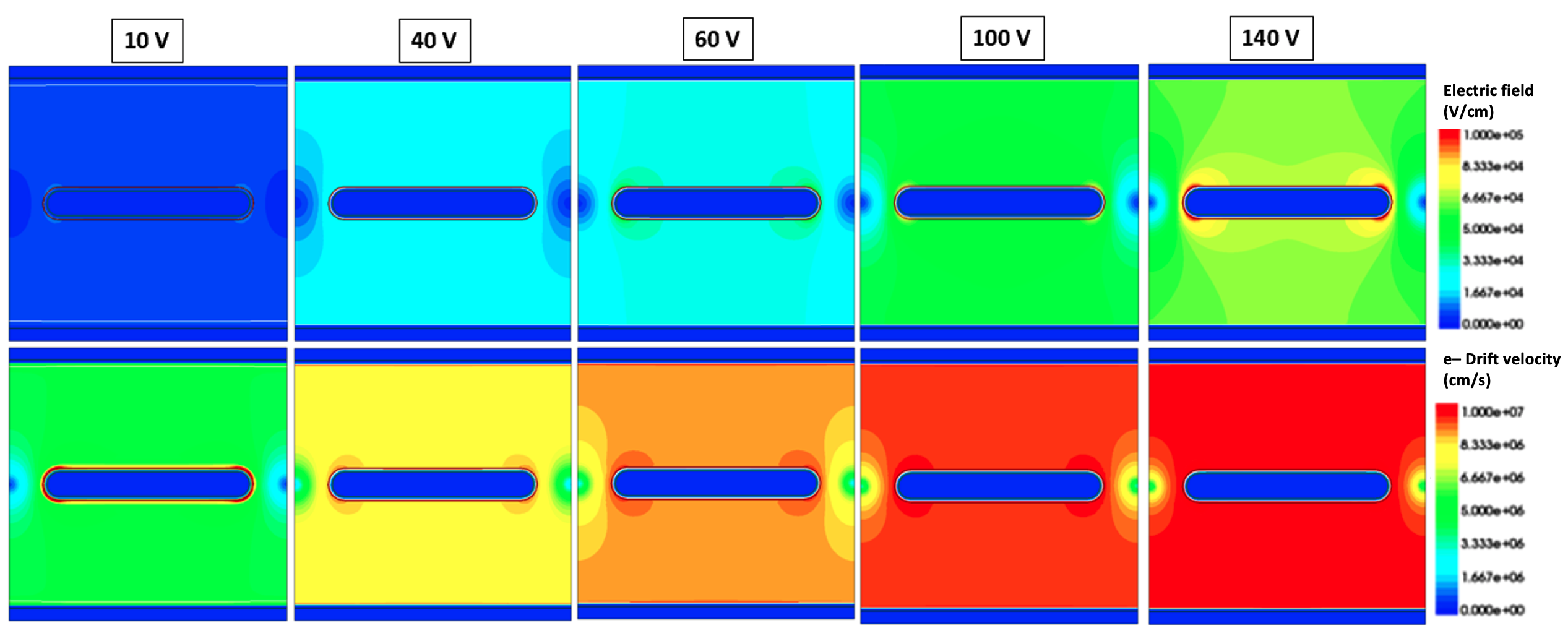}
    \caption{\footnotesize{Electric field and electron drift velocity with increasing bias voltage. At 100 V the electric field is sufficiently high to guarantee velocity saturation over the entire active area.}}
    \label{fig:driftfield}
    \end{figure}

As outlined in section~\ref{sec:sss_timing}, the basic criteria for timing optimisation consists in obtaining maximally uniform and high values of the \emph{i--let} contributions to the induced current signal
\begin{equation}
   i(t,\vec{r})= q \, \vec{E}_w \cdot \vec{v}_d.
    \label{eq:a1}
\end{equation}
Uniform sensor response is achieved by operating the sensor at sufficient high bias voltages so as to generate electric fields of magnitude greater than 10 kV/cm, needed to accelerate electrons and holes up to velocity saturation. In 3D silicon sensors relatively low bias voltages are sufficient to achieve the saturated velocity regime, as shown in Figure \ref{fig:driftfield}.

The second ingredient for fast sensor response is the weighting field $\vec{E}_w$, which couples the current induction over the entire active volume with the sensor geometry and, therefore, only depends on the electrode shape. The crucial role of the weighting field is further considered in the following subsection.
    
\subsection{2D geometry study: the Ramo maps}

The first step of sensor design is a preliminary study aimed at defining a limited group of geometries potentially suited for fast timing. This step is based on bi-dimensional quasi-stationary TCAD simulations (named 2D--TCAD). It consists in designing a horizontal cut of the pixel in different electrode geometries, according to its 3D symmetry. This 2D model of the device is simulated with increasing bias voltage and its relevant physics properties are determined at a temperature of 300 K and an inverse bias voltage of 100 V. This voltage is considered as a good reference to immediately exclude geometries in which drift velocity is still below the saturation regime in some areas. In this phase, different electrode geometries are designed, simulated and evaluated, featuring different combinations of square and hexagonal pixels with trench- or column-like electrodes. 

In the 2D--TCAD study, a new approach was developed to better identify electrode geometries for fast timing applications. The method, based on the Ramo theorem, is called \textit{Ramo map} and involves building a 2D map of the electron or hole \emph{i-let} values across the area. The Ramo map shows directly how strong and uniform a single carrier current induction is in a specific electrode geometry, allowing to identify critical spots and the potential single carrier output swing of the sensor. 
    
\begin{figure}[h!]
    \centering
    \includegraphics[width=0.7\textwidth]{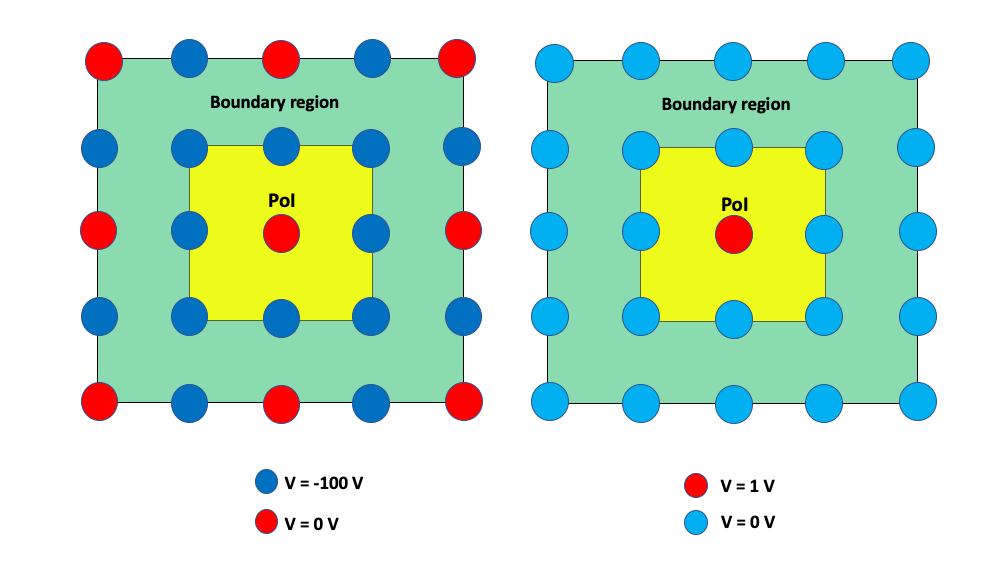}
    \caption{\footnotesize{Boundary conditions for a Ramo map: Left: boundary conditions for quasi stationary device simulation needed to compute drift velocity maps. Right: boundary conditions for the weighting field.}}
    \label{fig:RAMOsch}
    \end{figure}

\begin{figure}[h!]
    \centering
    \includegraphics[width=1.0\textwidth]{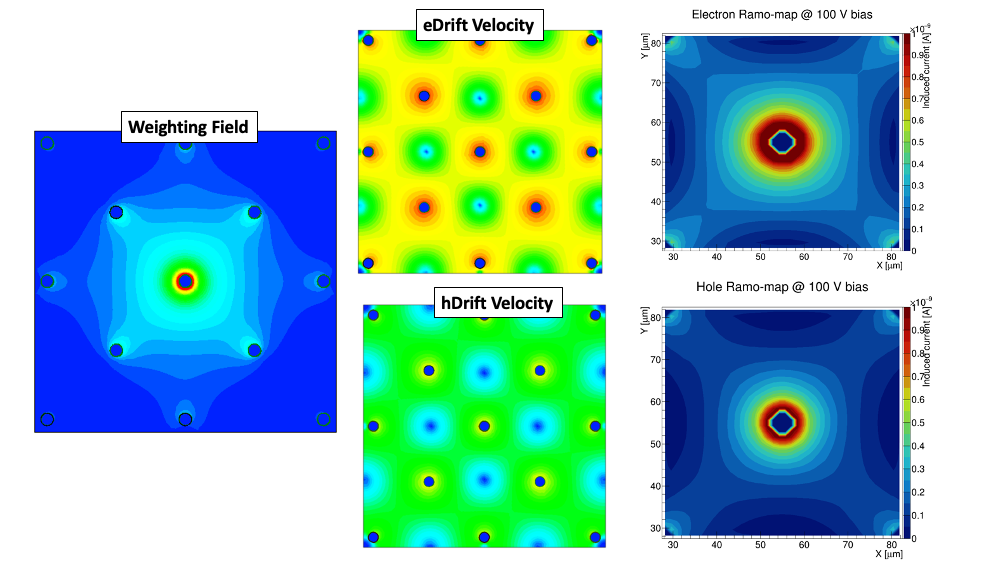}
    \caption{\footnotesize{Ramo map generation for a 5-column squared geometry from the calculation of the weighting field and carrier velocity maps.}}
    \label{fig:RamoMap}
    \end{figure}

A Ramo map is generated starting from the 2D model of an entire $3\times3$ pixel matrix (figure~\ref{fig:RAMOsch}). The innermost pixel is the pixel under study and is called \textit{Pixel of Interest} or PoI. The other boundary pixels are needed to better describe the weighting field of the PoI. A first quasi-stationary simulation is performed with normal operational voltages and is needed to compute the drift velocity maps of the device. TCAD quasi-stationary simulations consist in simulating the device by ramping one of the parameters which define its boundary conditions, usually the voltage. The simulation proceeds to calculate the physics of the device until the final value of the ramp is reached or the solution diverges. The physical information of the sensor at specific boundary conditions, for example at --100 V bias voltage, are extracted and used for the definition of the Ramo-map \cite{tcad}. A second quasi-stationary simulation is used to compute the weighting field. In this case the boundary conditions have a potential V = 0 on all the electrodes with the exception of the  readout electrode of the PoI, which is set to V = 1. Weighting field and drift velocity maps are then scalar-multiplied point by point over the entire area according to the Ramo theorem (figure~\ref{fig:RamoMap}).   
 
\begin{figure}[h!]
    \centering
    \includegraphics[width=0.85\textwidth]{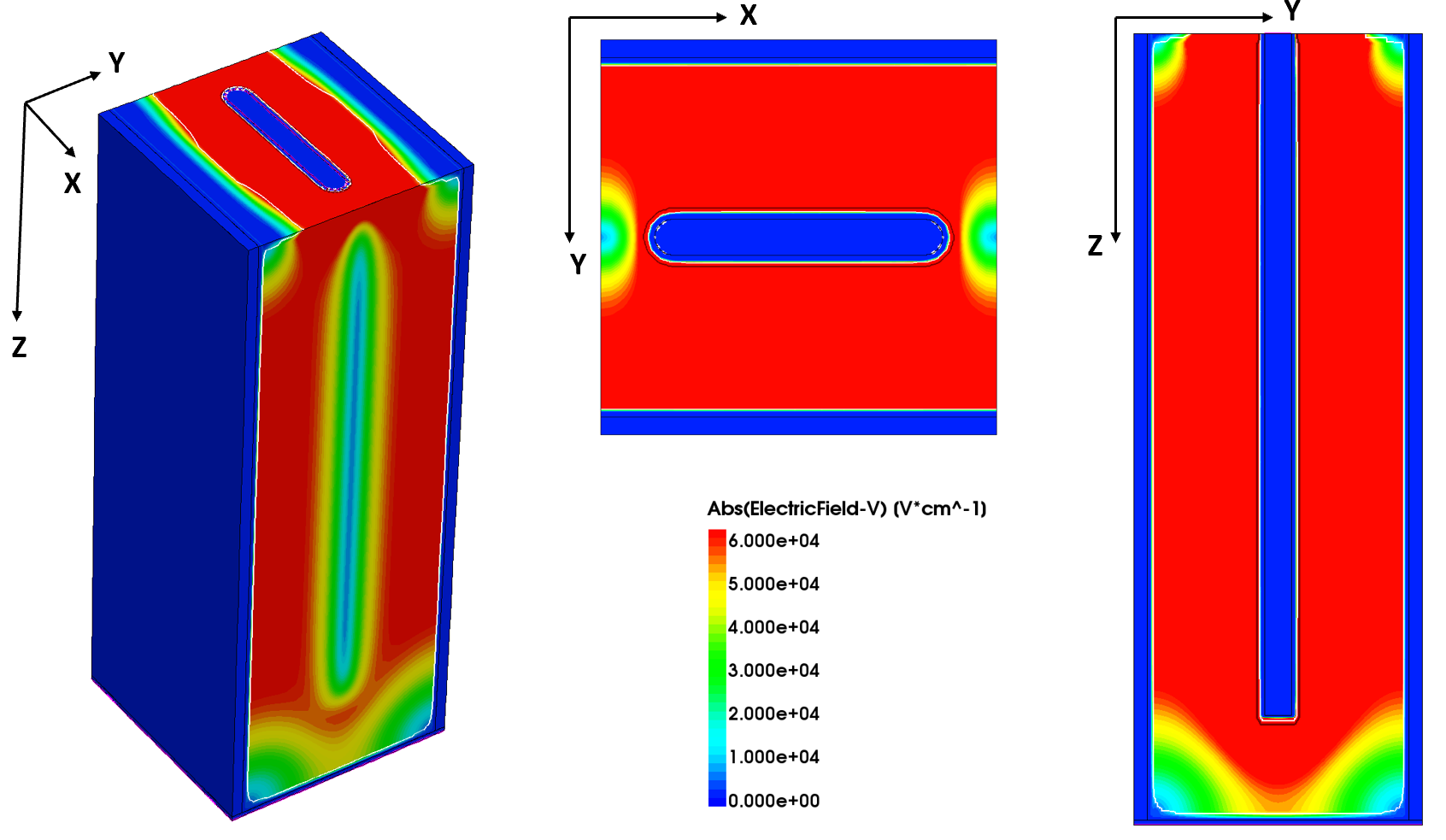}
    \caption{\footnotesize{Full 3D-device simulation of the electric field of the TIMESPOT parallel trench pixel at $V_{bias} = -100 V$.}}
    \label{fig:3Drender}
    \end{figure}

The Ramo map gives a first quantitative estimate of timing performance of a given 3D sensor geometry. An ideal Ramo map has uniform and high \emph{i-let} values. This design and simulation approach based on 2D-device simulation allows to explore in a more efficient way the potential timing performances of multiple electrode geometries, avoiding to use more detailed but also time consuming full-3D device simulation. Full 3D-device simulation is dedicated only to those geometries which were selected during this first phase of the study (as can be seen in figure~\ref{fig:3Drender}). 

\begin{figure}[h!]
    \centering
    \includegraphics[width=0.9\textwidth]{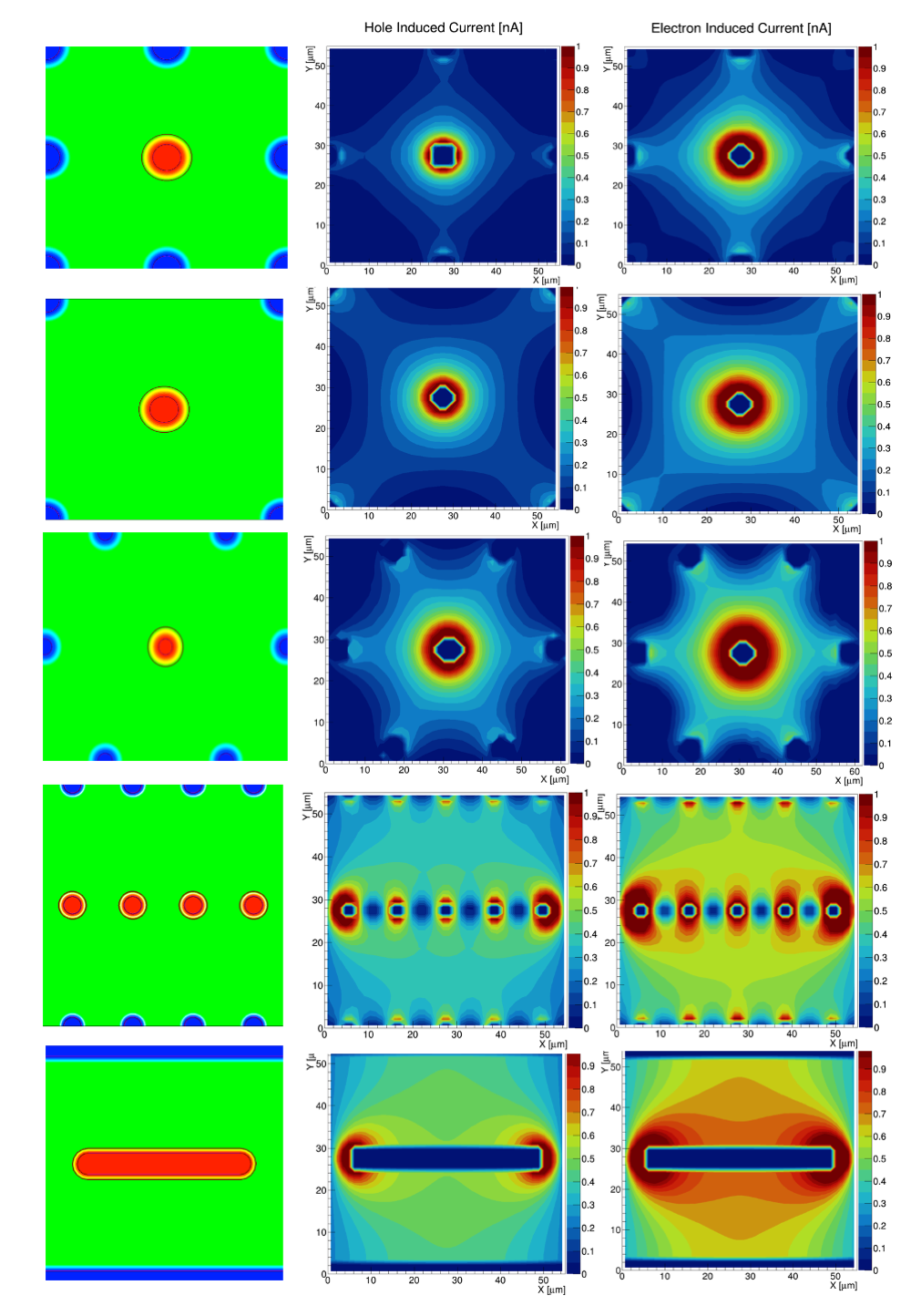}
    \caption{\footnotesize{Ramo map comparison of five different electrode geometries. From left to right: doping profiles (red: n++ doping, blue: p++ doping), hole Ramo map, electron Ramo map. From top to bottom: squared 9-columns, squared 5-columns, hexagonal 7-columns, parallel 12-columns, parallel trench.}}
    \label{fig:RamoMapComp}
    \end{figure}

Figure~\ref{fig:RamoMapComp} shows a comparison of three different electrode geometries by means of their Ramo maps. Among all different geometries, those with a parallel configuration of their electrodes have the highest current induction with the smallest output current swing. The parallel trench geometry shows the best features. More classic geometries, like the 5-column pixel device, used in the ATLAS-IBL~\cite{ATLAS-IBL, ATLAS-IBL2} or considered for the new ATLAS Inner Tracker~\cite{ATLAS-ITk}, have \emph{i-let} values which are lower in most parts of the sensor area. Moreover, they are less uniform and consequently give a much larger output current swing.   

 \begin{figure}[h!]
    \centering
    \includegraphics[width=0.85\textwidth]{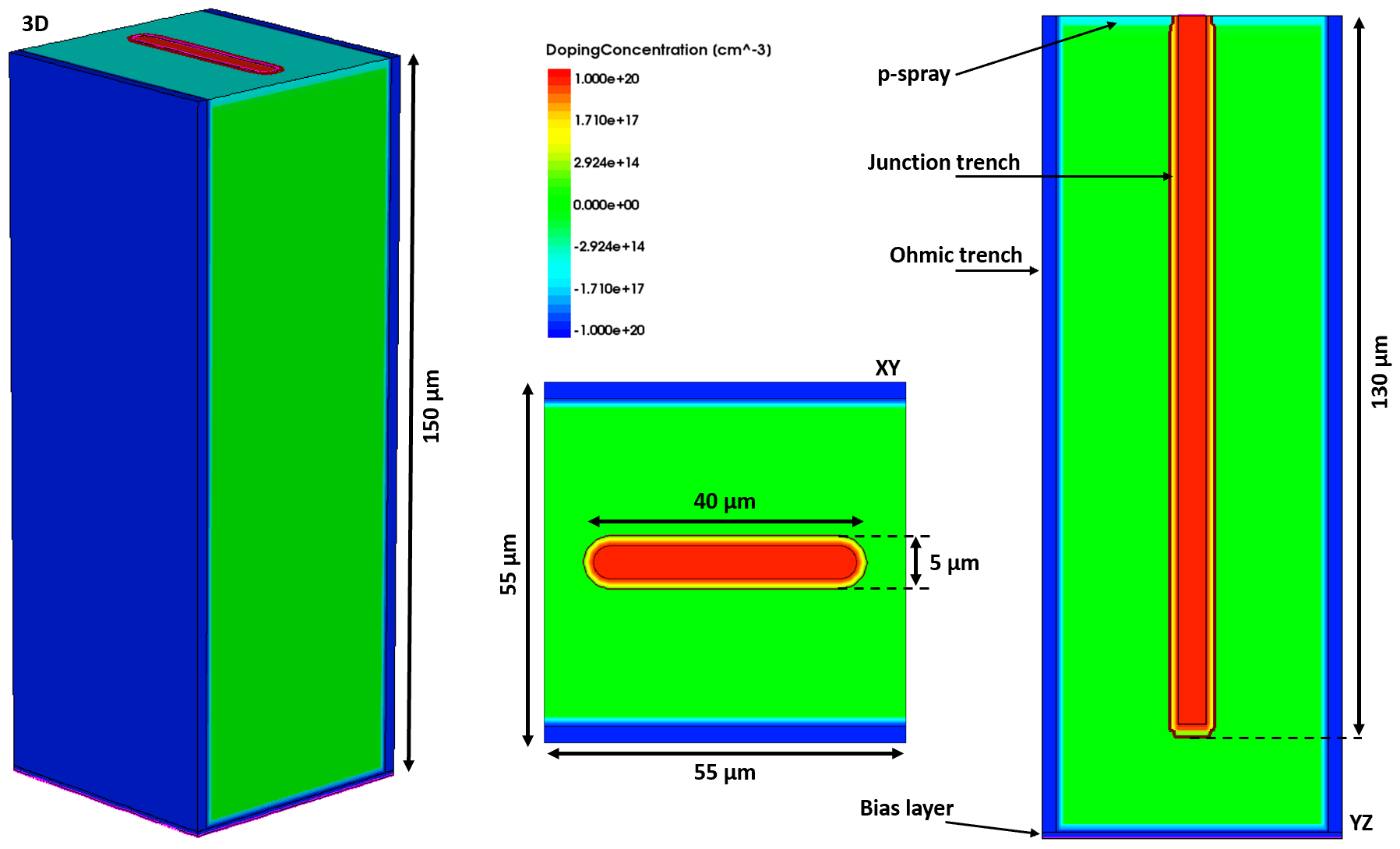}
    \caption{\footnotesize{3D rendering of the TIMESPOT parallel trench device showing doping concentration and internal structure. Radiation hardness along the surface is increased using a p-spay layer. The bottom of the sensor presents the residual layer of the support wafer which is used to provide the bias voltage to the ohmic electrodes.}}
    \label{fig:3dtrench}
    \end{figure}  
    
\subsection{3D design}
After geometry selection, a complete 3D model is designed and simulated,  adding more details of the fabrication process such as the 20$\mu$m shorter junction electrode with respect to the ohmic ones, a p-spray layer on top of the pixel in order to increase radiation hardness against superficial damage and a thin layer of low resistive and p++ doped silicon layer from the support wafer which is left after the fabrication process, providing bias from the bottom (figure~\ref{fig:3dtrench}). Detailed simulations of the model are performed in order to have a more detailed overview of the real device, estimate its capacitance using small AC signal analysis and generating the physics, using quasi stationary TCAD simulations, needed to perform transient simulation using the TCoDe fast transient simulator. For such a purpose, the complete 3D model outputs 3D maps of the electric field (figure~\ref{fig:3Drender}), carrier mobilities and weighting field are produced. The structure and operation of the TCoDe simulator itself is described in the next section.

\section{Simulation of carrier dynamics and signal generation: the TCoDe software}
\label{sec:3D_optim}
TCoDe~\cite{tcode} (TIMESPOT Code for Detector simulation) is a $C^{++}14$ --compliant application to simulate the response of solid state sensors in massively parallel platforms on Linux systems. TCoDe is implemented on top of Hydra~\cite{hydra} and as such, it can run on OpenMP \cite{OPENMP}, CUDA \cite{CUDA} and TBB \cite{TBB} compatible devices. 

TCoDe uses external 3D maps of electric fields, carrier mobilities, weighting field and energy deposit to simulate the induced-current response of solid state sensors. The motion of the individual carriers produced in the initial deposit is determined with a $4^{th}$ order Runge--Kutta algorithm using electric field and mobility maps and assuming that the carriers always move at drift velocity. At each time interval the current induced in the electrode is calculated from the carrier velocity using the corresponding weighting field, according to the Shockley--Ramo theorem. The output is stored to ROOT~\cite{root} files, several level of output detail are available, from the simple current vs time plot to the complete information about the position of the carriers at each time step.

\subsection{Energy deposits}
TCoDe is able to generate its own energy deposits or import them from other tools like GEANT4 \cite{GEANT4}. The deposit is determined by its energy, $E_{dep}$, released along a path of length $l$, identified by the start and end point of the ionization path ($(X_i;Y_i;Z_i)$ and $(X_f;Y_f;Z_f)$), the energy deposit $E_{dep}$, the dispersion of the charges at the beginning $\alpha_i$ and end $\alpha_f$ of the path. Distinction between primary and secondary contributions, such as $\delta$ rays is also possible. In this latter case, the deposit is composed by several straight segments.
The carriers distribution is created by converting these straight energy deposits into an equivalent number of electron-hole pairs. This operation is performed by dividing the energy deposit by the average energy needed to generate an electron hole pair which, in case of silicon, is equal to 3.6 eV. At the end of the process there will be a number of point-like carriers, $N_{max}$ , distributed in space.

\subsection{Sensor geometry and physics maps}
TCoDe is sensor-agnostic, in the sense that there is no assumption on the shape or its features. The sensor is entirely defined by physical quantities, or maps, provided across the sensor volume: electric field vector, electron and hole mobilities, weighting field.
To be more efficient in the calculations these quantities have to be provided in points on a 3D grid, not necessarily a regular one, to allow a fast lookup via binary search, depending on the carrier position. The grid does not have to be the same for all physical quantities, although if this is true, the calculation will be faster. If the physics maps are not provided in this format, they will have to be converted through interpolation.
In general a carrier will not sit exactly on one of these grid points, but it will be in a position surrounded by 8 of them so that the physical quantities can easily be calculated through a simple linear interpolation.
Since physics maps can be very large depending on the sensor, some optimisation is required, for example by using a coarser grid in regions where the quantities vary slowly. However the size of the grids (and therefore their level of refinement) has to be decided depending on the hardware available, there is no limitation coming from the TCoDe software. Physics maps used in this paper have roughly 1M points and are a few giga-bytes large.

\subsection{Carrier motion}
The motion of each carrier is treated separately and interaction among them is neglected. This can generally be assumed for 3D sensors since there is no multiplication and therefore the charge density is sufficiently low. 
Charge drift is computed by solving the following differential equation which describes charge motion in semiconductors under effects of an electric field:
\begin{equation}
\dfrac{{\delta}\vec{r}_{d}}{{\delta}t} = {\mu}\vec{E}
\label{eq:5.10}
\end{equation}
As integration method the Runge--Kutta $4^{th}$ order algorithm was implemented. The algorithm computes for the time step $N$ the drift motion step $\vec{r_{d}}_{N+1}$.
\begin{equation}
\vec{r_{d}}_{N+1} = \vec{r_{d}}_{N} + \frac{{\Delta}t}{6}(k_1 + 2(k_2 + k_3) + k_4)
\label{eq:5.11}
\end{equation}
With $k_{1,2,3,4}$ the Runge--Kutta parameters, which are:
\begin{equation}
\begin{cases} 
\vec{k_1} = \mu_{e,h}(t_n,\vec{r_{d}}_{N})\vec{E}(t_n,\vec{r_{d}}_{N})\\
\vec{k_2} = \mu_{e,h}(t_n + \frac{t_{step}}{2} , \vec{r_{d}}_{N} + \frac{1}{2}\vec{k_1}t_{step})\vec{E}(t_n + \frac{t_{step}}{2} , \vec{r_{d}}_{N} + \frac{1}{2}\vec{k_1}t_{step})\\
\vec{k_3} = \mu_{e,h}(t_n + \frac{t_{step}}{2} , \vec{r_{d}}_{N} + \frac{1}{2}\vec{k_2}t_{step})\vec{E}(t_n + \frac{t_{step}}{2} , \vec{r_{d}}_{N} + \frac{1}{2}\vec{k_2}t_{step})\\
\vec{k_4} = \mu_{e,h}(t_n + \frac{t_{step}}{2} , \vec{r_{d}}_{N} + \frac{1}{2}\vec{k_1}t_{step})\vec{E}(t_n + h , \vec{r_{d}}_{N} + \vec{k_1}t_{step}) 
\end{cases}
\label{eq:5.12}
\end{equation}
The electric field $\vec{E}$ and mobility $\mu_{e,h}$ are extracted or interpolated from the input physics maps.

\subsection{Carrier diffusion for thermal effects}
Finite temperature adds a random component to the carrier motion. This is quantified by the diffusion coefficient $D$ which depends on the temperature of the silicon
\begin{equation}
D = \dfrac{k_BT}{e}\mu_{e,h},
\label{eq:5.6}
\end{equation}
where $k_B$ is the Boltzmann constant and $\mu_{e(h)}$ are the electron (hole) mobility which is directly taken from the mobility maps. For a time step ${\Delta}t$ the path traveled by the charge is equal to
\begin{equation}
\sigma_{diff} = \sqrt{2D{\Delta}t} 
\label{eq:5.7}
\end{equation}
Diffusion direction is a completely random process and for every time step its direction and distance varies randomly. In order to include this aspect in the simulation, $\sigma_{diff}$ is computed as the variance of a random number generator with Gaussian distribution with mean value 0. This allows to compute for every ${\Delta}t$ a random drift distance within the desired distribution. The orientation of the diffusion step is also randomly computed by rotating the vector of length $\sigma_{diff}$ around the origin with an angle $\theta$ and $\phi$. 
\begin{equation}
{\delta}\vec{r}_{diff}=\begin{pmatrix}
X_{diff} \\ 
Y_{diff} \\ 
Z_{diff}
\end{pmatrix} = \begin{pmatrix}{}
cos(\phi)sin{\theta} & -sin(\phi) & -cos(\phi)sin{\theta} \\ 
sin(\phi)cos{\theta} & cos(\phi) & -sin(\phi)cos{\theta} \\ 
sin(\theta) & 0 & cos{\theta}
\end{pmatrix}\times\begin{pmatrix}
\sigma_{diff} \\
\sigma_{diff} \\
\sigma_{diff} 
\end{pmatrix}
\label{eq:5.8}
\end{equation}
With $\theta$ and $\phi$ randomly generated angles between $[-\pi,\pi]$ and $[0,2\pi]$ respectively. $\delta\vec{r}_{diff}$ is simply added to the carrier motion due to the electric field.
At the end of the time step, $\vec{r_{d}}_{N+1}$ and ${{\delta}\vec{r}_{diff}}_{N+1}$ are added together to obtain the total distance travelled by the charge:
\begin{equation}
\vec{r}_{N+1} = \vec{r}_{d_{N+1}} + \vec{r}_{diff_{N+1}}
\label{eq:5.13}
\end{equation}

\subsection{Induced current}
The current induced on the electrode  is calculated by applying the Ramo theorem  
\begin{equation}
i_{e,h} = q_{e,h}(v_{x_{e,h}}E_{w_x} + v_{y_{e,h}}E_{w_y} + v_{z_{e,h}}E_{w_z}),
\label{eq:5.14}
\end{equation}
where $E_{w_{x,y,z}}$ is the weighting field at the position of the carrier. The contribution of every single charge is added at the end of each time step and saved on the output file. In default settings the output includes the time step, the total induced current in time and integrated charge in time. It is also possible to save the contributions of the electrons and holes separately, as well as the charge generated by primary and secondary particles. In this way, it is possible to analyze the contribution of electrons and holes separately or the contribution from deposits generated by primary or secondary particles. Moreover, it is possible to save the drift path of every single charge and use the data to visualise the entire process.

\subsection{Multi-thread implementation}
The characteristics of TCoDe is to have the ability to perform very detailed simulations, by following each carrier individually, but at the same time being extremely fast thanks to its multi-thread implementation. Parallelisation is used whenever possible in the code, however most of the performance comes from treating all the carriers simultaneously so that each time step bunches of carriers can be processed in parallel. 
TCoDe can parallelise both in CPU (OMP, TBB), or NVIDIA GPU (CUDA). In general, it supports all back-ends supported by Hydra, on top of which TCoDe is built.
It is clear that the performance depends strictly on the number of parallel threads and therefore on the hardware available. Moreover, all calculations are performed in double precision. As it is now, TCoDe cannot simulate sensors where charge multiplication occurs. This is because if the number of carriers changes, then also the number of threads should. There are ways to overcome this issue, for example by grouping carriers, but this is not yet fully implemented.
Given a set of physics maps and a certain hardware, there are two parameters of the simulation that allow to reach the best compromise between detail and performance: these are the number of carriers and the time step. The user has to find the best compromise between these two. As an extra handles, charges can be treated in groups, so that the effective number of threads is reduced by the size of the groups.

Typical performances on consumer desktop and laptop computers are reported on \tablename~\ref{tab::performance}.

\begin{table}[ht]
\centering
\begin{tabular}{ |c|c|c| }
 \hline
 & \multicolumn{2}{|c|}{\textbf{Simulation time [sec]}} \\
 \hline
 \textbf{Backend} & \textbf{Desktop PC} & \textbf{Laptop} \\
 \hline
 Single thread & 24.5 & 29 \\
 \hline
 TBB & 5.3 & 6.8 \\
\hline 
 OMP & 5.6 & 8.4 \\
\hline
 CUDA & 1.3 & 3.5 \\
 \hline
 
\end{tabular}
\caption{Calculation time for a simulated MIP deposit of 12000 hole-electron pairs in the parallel trench sensor. A total of 2000 1-ps time steps are executed.The timing is shown for a desktop pc (Intel Core i7-6700K processor, 16 GB RAM and an NVIDIA GeForce GTX 1080) and a laptop (Intel Core i7-7700HQ, 16 GB RAM and an NVIDIA GeForce GTX 1050).}
\label{tab::performance}
\end{table}

It is noted that since double precision is required, only double precision cores are used in GPUs. Therefore, performances are expected to improve significantly on professional GPU cards.

\section{Timing behaviour of 3D silicon sensors}
\label{sec:time_behave}

In the present section, two examples of design  and data analysis methodologies using the TCoDe simulator are shown. 
In the first case (subsection~\ref{subsec:geomcomp}), TCoDe is used on full size 3D pixel models to perform a detailed study concerning the timing performance foreseen for 3D sensors according to their geometries.
In the second case (subsection~\ref{subsec:performancestudy}) TCoDe shows its potential in studying and understanding the timing behavior of sensors already tested in the laboratory.

\subsection{Comparison between timing performance of different 3D structures}
\label{subsec:geomcomp}

\begin{figure}[h!]
    \centering
    \includegraphics[width=0.9\textwidth]{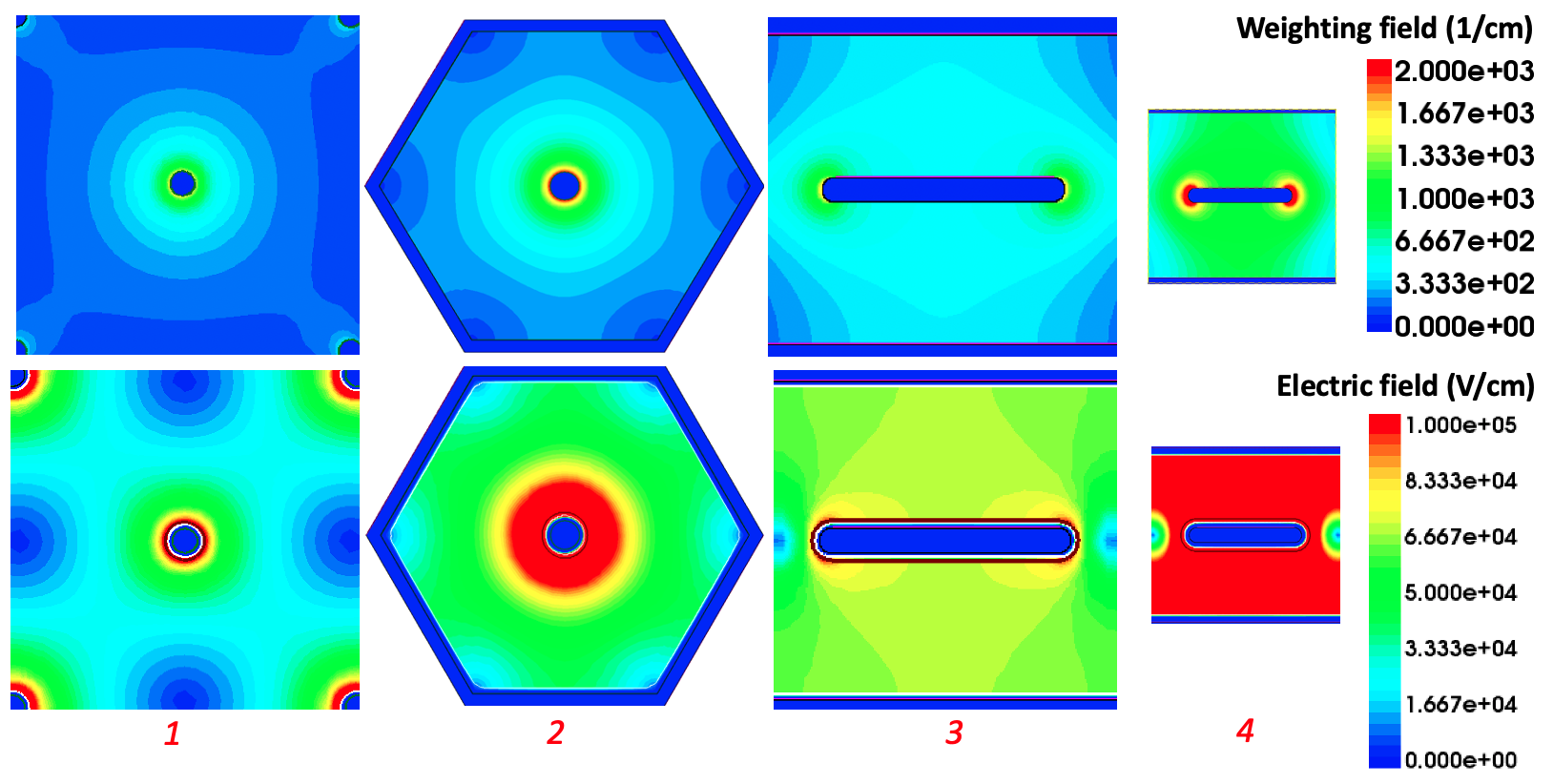}
    \caption{\footnotesize{Transverse cuts of 3D pixels with different geometries. Weighting field maps (top row) and Electric field maps (bottom row) at $V_{bias} = -150V$. Geometries 1 to 3 have pitch = 55$\mu$m, while 4 has pitch = 25$\mu$m. TCAD simulations.}
    }
    \label{fig:Geom4comp}
\end{figure}

 The TCoDe simulator is used here to compare the timing performance of four different 3D geometries. As performance indicator we will use the Charge Collection Time (CCT) distribution. The CCT is defined as the time needed for all the carriers generated by a given track to reach their respective electrodes, that is for the induced current signal to return to zero. As demonstrated in~\cite{LaiCossu}, the CCT distribution can be strictly correlated to the final time resolution of the detecting system in such a way to create a direct connection between the standard deviation $\sigma_{t_c}$ of the CCT distribution and the one of the time of the arrival distribution $\sigma_t$. The four geometries under comparison are shown in figure~\ref{fig:Geom4comp}.
 
\begin{figure}[h!]
    \centering
    \includegraphics[width=1.0\textwidth]{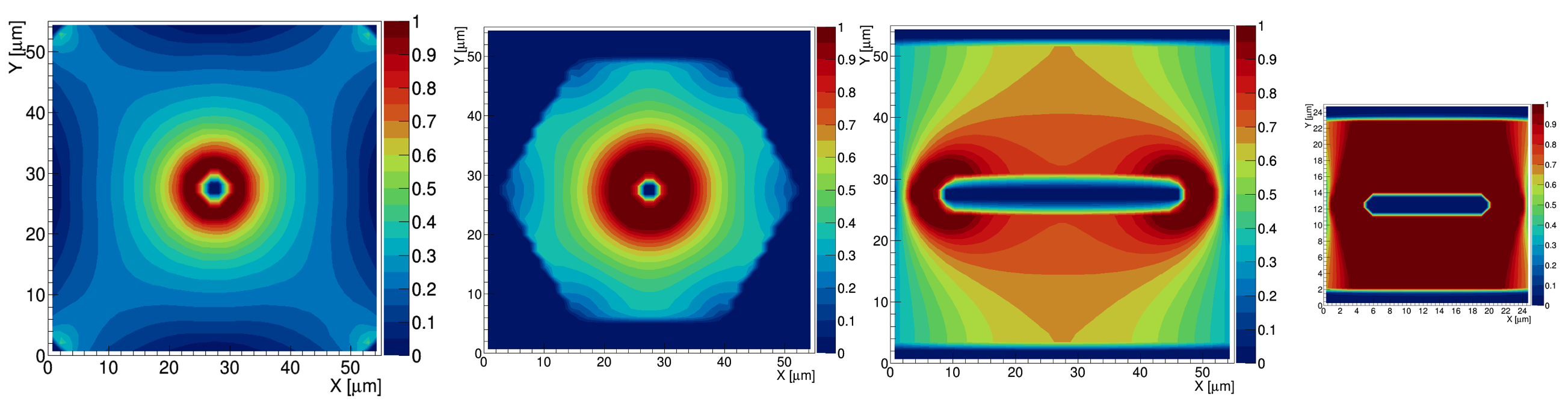}
    \caption{\footnotesize{Electron Ramo map comparison for the four 3D geometries of figure~\ref{fig:Geom4comp}. }
    }
    \label{fig:RamoMap4Comp}
\end{figure}

The Ramo maps of the four geometries are reported in figure~\ref{fig:RamoMap4Comp}. It is noted that the trench geometries provide more uniform and higher \emph{i--let} values. This fact suggests that the 3D-trench geometry will have higher induced current signals, shorter signal time, faster charge collection and, consequently, better time resolution.

\begin{figure}[h!]
    \centering
    \includegraphics[width=1.0\textwidth]{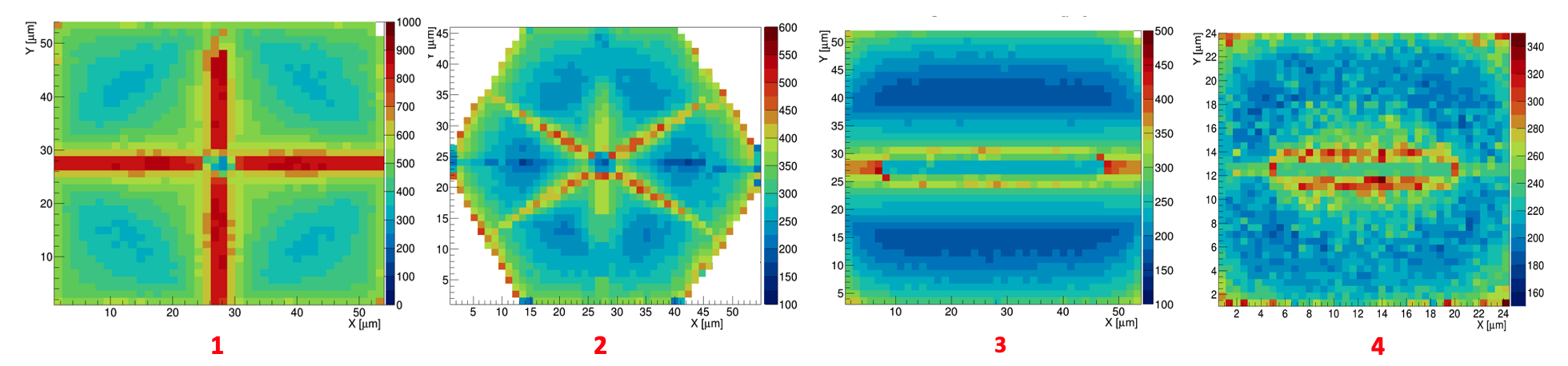}
    \caption{\footnotesize{2D-maps TCoDe-simulated Charge Collection Times [ps] for four different geometries (1-4) reported in figure~\ref{fig:Geom4comp}. Please note the different scales in the color code. }
    }
    \label{fig:CCTmaps}
\end{figure}

To explore the time response of the different sensor geometries, a TCoDe simulation is performed. The simulation is obtained by analysing the effect of about 2\,500 MIP perpendicular tracks, scanning completely the area of each type of sensor in steps of $\approx$ 1$\mu$m. The result is illustrated in figure~\ref{fig:CCTmaps}, which shows the simulated CCT for each point of the pixel area. Furthermore, the CCT maps allows an immediate and detailed check about the \emph{weak spots} inside the pixel volume.

\begin{figure}[h!]
    \centering
    \includegraphics[width=0.75\textwidth]{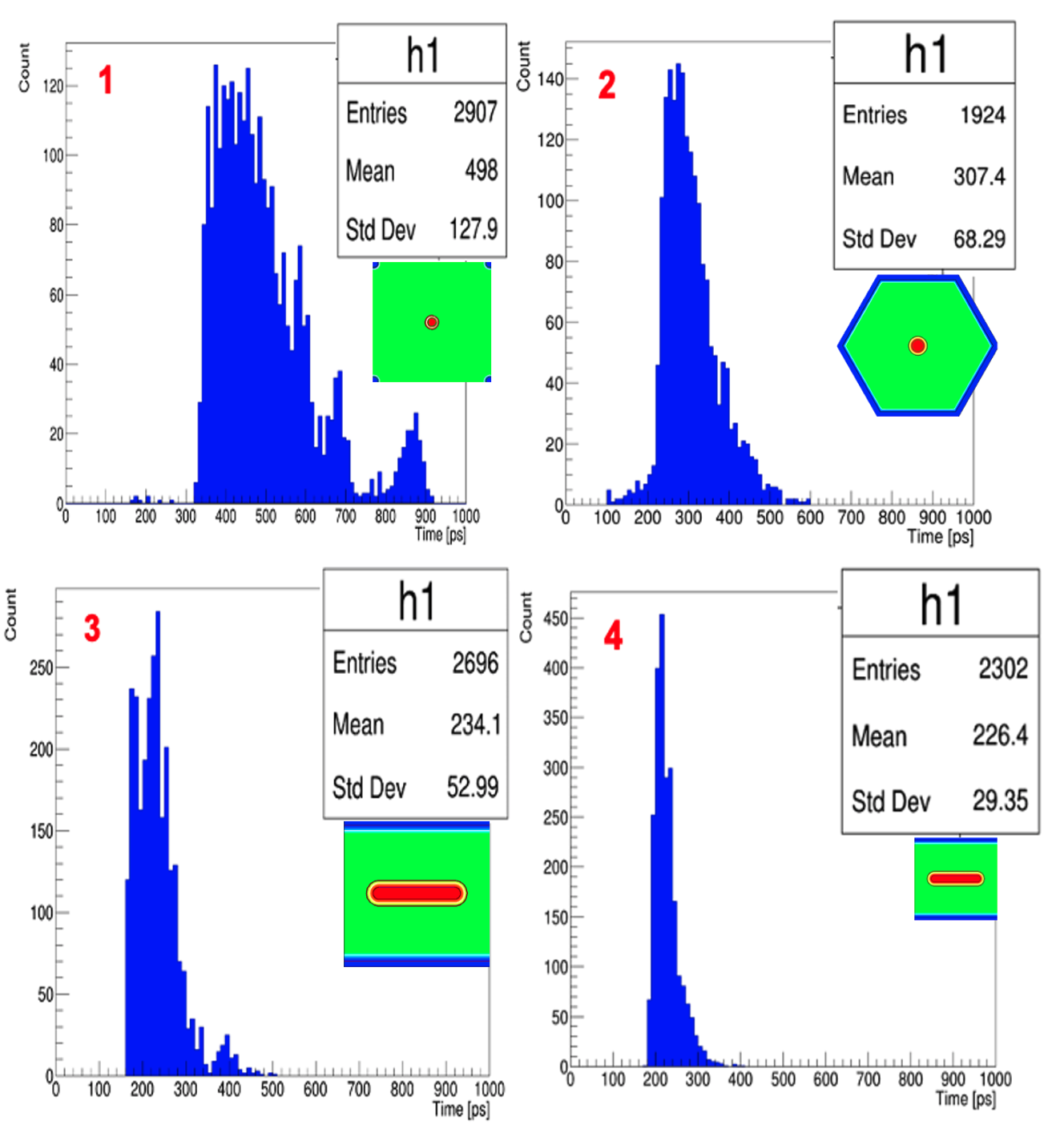}
    \caption{\footnotesize{TCoDe-simulated distributions of Charge Collection Times [ps] for the geometries (1-4) of figure~\ref{fig:Geom4comp} to~\ref{fig:CCTmaps} (see insets).}
    }
    \label{fig:CCTdistr}
\end{figure}

The next step of this analysis is getting the CCT distributions which sum-up the internal contributions of the whole pixel. These are given in figure~\ref{fig:CCTdistr}. From the obtained distributions two important parameters che be extracted: the mean CCT or \emph{time centroid}~\cite{Riegler} of the pixel, or $t_c$, and the standard deviation, or $\sigma_{t_c}$, which can give useful predictions on the final time resolution of the system~\cite{LaiCossu}:
\begin{equation}
   \sigma_t = P\sigma_{t_c},
\end{equation}
where $P$ is called \emph{timing propagation coefficient} and is a function of the front-end properties.
Since such analysis goes beyond the scope of the present work, here we can limit to observe the effectiveness of the TCoDe analysis in describing and predicting the time behaviour of sensors and in particular of the 3D technology sensors. Moving from the 3D-column geometry (number 1) to the parallel trench geometry (number 3) provides more than a factor two reduction in $\sigma_{t_c}$. Pitch reduction (geometry number 4) can improve the effect almost by an additional factor two.

\subsection{Study of measured timing performance}
\label{subsec:performancestudy}

\begin{figure}[h!]
	\centering
	\includegraphics[width=1.0\linewidth]{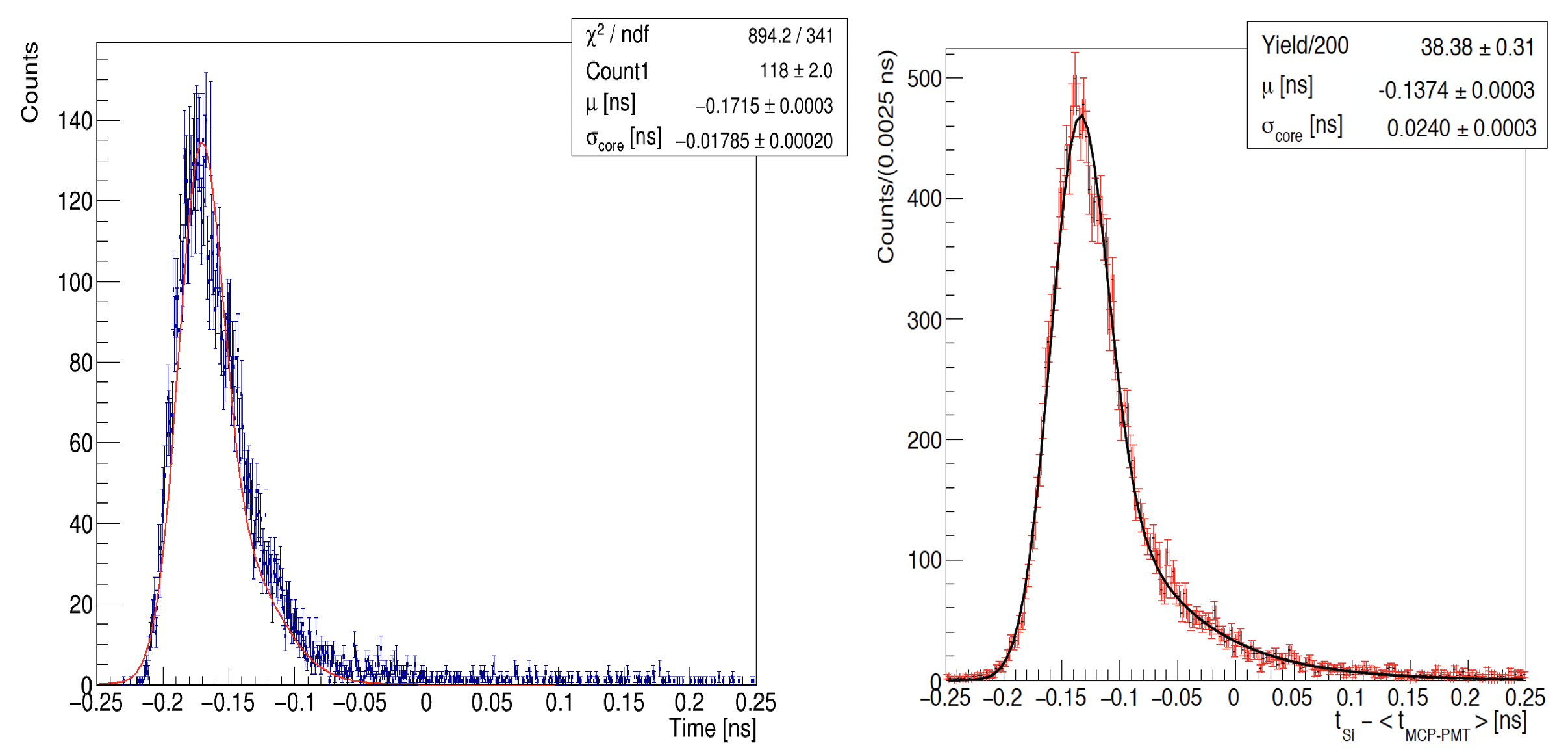}
	\caption{Simulated (left) and measured (right) distributions of ToA for a 3D-trench sensor.}
	\label{fig:TOAdist}
\end{figure}

As shown in the previous section, TCoDe can be used to realise virtual and controlled experiments on sensors, which can allow an almost one-to-one comparison with measurements. Starting from the detailed description of the current signals as induced at the electrodes \emph{i(t)} and considering the front-end transfer function \emph{H(t)}, it is possible to obtain the front-end response via the convolution
\begin{equation}
    v_{out}(t) = i(t) * H(t).
\end{equation}
The distribution of the Time of Arrival of the $v_{out}(t)$ signals can then be built by applying a suitable discrimination algorithm.
Figure~\ref{fig:TOAdist} (left) gives the simulated distributions of the Time of Arrival (ToA) of the $ v_{out}(t)$ signals.
The ToA of each signal was obtained by applying a constant fraction discriminator algorithm with a threshold set at 35\% of the signal amplitude. The simulated ToA distribution is compared in figure~\ref{fig:TOAdist} to a ToA distribution directly coming from test-beam measurements~\cite{PSIres}. The measured ToA distribution includes the contribution of the time references (two devices based on Micro-Channel-Plates~\cite{PSIres}), which is summed in quadrature. The simulated ToA considers the time delay between the read-out output and the simulation start, established by the charge deposit inside the sensor. This yields a significantly higher $\sigma_{core}$ value for the measured distribution. When the time reference contribution is subtracted, the value $\sigma_{core}=(20.6\pm 0.4)$~ps is obtained~\cite{PSIres}. The simulated behavior already well reproduces the shape of the measured one. To obtain an accurate quantitative comparison, a more precise modeling of the front-end and of its noise behaviour is to be applied.
    
\begin{figure}[h!]
	\centering
	\includegraphics[width=1.0\linewidth]{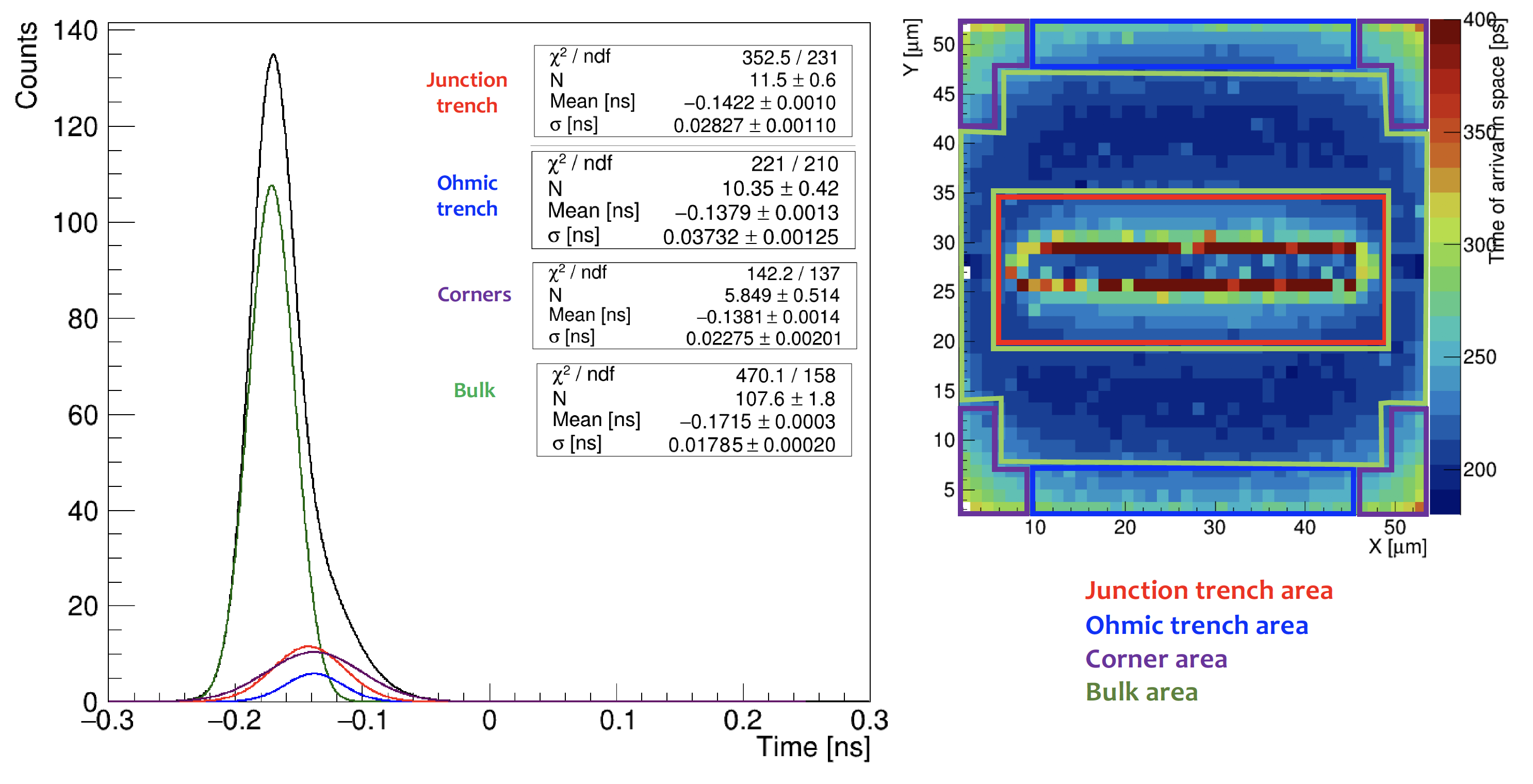}
	\caption{Decomposition of the ToA distribution of figure~\ref{fig:TOAdist} (left) according to different pixel areas (right).}
	\label{fig:AreasDist}
\end{figure}

TCoDe can be used to accomplish a dedicated study of the ToA distribution structure of the type shown in figure~\ref{fig:TOAdist}. The distribution can be analysed by separating the contributions of the single time distribution per different pixel areas. Figure~\ref{fig:AreasDist} reports such analysis and indicates the correspondences between the single sub-distributions and the associated pixel sub-areas. The plot points out the \emph{weak} sub-areas which slow down the global performance. This clarifies the origin of the tail in the distributions of figure~\ref{fig:TOAdist} and can suggest suitable modifications in the pixel geometry to further optimise its time resolution.

\section{Conclusions}
\label{sec:conclusion}

The present work describes the structure and functionalities of the TCoDe simulator. It is released with a \emph{GPLv3} license and its open-source code can be found and downloaded at the link indicated in~\cite{tcode}. Some examples have been given about the TCoDe effectiveness in the design of 3D silicon sensors, when coupled to other tools for electric field description, as well as in the analysis and interpretation of data from measurements. Although its development is mainly dedicated to 3D silicon sensor modeling, TCoDe is not technology-specific and is easily transportable to the modeling of the carrier transportation mechanisms of any solid state sensor. In particular, its multi-thread core makes perfectly feasible to realize full-scale virtual experiments in reasonable time and on medium-level machines (see Table~\ref{tab::performance}), where other methods are simply impractical solutions even when used on high-end computing infrastructures. 

\acknowledgments

The authors are grateful to Gian Matteo Cossu for his help in providing the front-end transfer function used in figures~\ref{fig:TOAdist} and~\ref{fig:AreasDist}. 

This work was funded and supported by the Fifth Scientific Commission (CSN5) of the Italian National Institute for Nuclear Physics (INFN), within the Project TimeSPOT. It has also received funding from the ATTRACT project, funded by
the EC under Grant Agreement 777222 and by the ATTRACT-EU initiative, INSTANT project.

\bibliographystyle{ieeetr}  
\bibliography{ArtTCoDe}  

\end{document}